**A method for estimating spatial resolution of real image in the Fourier domain**

Ryuta Mizutani*, Rino Saiga*, Susumu Takekoshi†, Chie Inomoto‡, Naoya Nakamura‡, Masanari Itokawa§, Makoto Arai§, Kenichi Oshima§, Akihisa Takeuchi‖, Kentaro Uesugi‖, Yasuko Terada‖, and Yoshio Suzuki‖

*Department of Applied Biochemistry, School of Engineering, Tokai University, Kitakaname 4-1-1, Hiratsuka, Kanagawa 259-1292, Japan
†Department of Cell Biology, Division of Host Defense Mechanism, Tokai University School of Medicine, Shimokasuya 143, Isehara, Kanagawa 259-1193, Japan.
‡Department of Pathology, Tokai University School of Medicine, Shimokasuya 143, Isehara, Kanagawa 259-1193, Japan
§Tokyo Metropolitan Institute of Medical Science, Kamikitazawa 2-1-6, Setagaya, Tokyo 156-8506, Japan
‖Japan Synchrotron Radiation Research Institute (JASRI/SPring-8), Kouto 1-1-1, Sayo, Hyogo 679-5198, Japan

Correspondence to: Ryuta Mizutani, Department of Applied Biochemistry, Tokai University, Hiratsuka, Kanagawa 259-1292, Japan. Tel: +81-463-58-1211; fax: +81-463-50-2506; e-mail: ryuta@tokai-u.jp

**Abstract**

Spatial resolution is a fundamental parameter in structural sciences. In crystallography, the resolution is determined from the detection limit of high-angle diffraction in reciprocal space. In electron microscopy, correlation in the Fourier domain is used for estimating the resolution. In this paper, we report a method for estimating the spatial resolution of real images from a logarithmic intensity plot in the Fourier domain. The logarithmic intensity plots of test images indicated that the full width at half maximum of a Gaussian point-spread function can be estimated from the images. The spatial resolution of imaging X-ray microtomography using Fresnel zone-plate optics was also estimated with this method. A cross section of a test object visualized with the imaging microtomography indicated that square-wave patterns up to 120-nm pitch were resolved. The logarithmic intensity plot was calculated from a tomographic cross section of brain tissue. The full width at half maximum of the point spread function estimated from the plot coincided with the resolution determined from the test object. These results indicated that the logarithmic intensity plot in the Fourier domain provides an alternative measure of the spatial resolution without explicitly defining a noise criterion.

**Second abstract**

Spatial resolution is a fundamental parameter in structural sciences. In this paper, we report a method for estimating the resolution from a logarithmic intensity plot in Fourier space. The obtained results indicated that the logarithmic plot provides an alternative measure of the resolution without defining a noise criterion.

**Introduction**

Spatial resolution is a fundamental parameter in structural sciences. Although the frequency responses of optical systems have been studied analytically and used to estimate the resolution of imaging detectors (Hopkins 1955; Koch et al., 1998), the resolution of real images depends on a number of experimental factors including mechanical drift and sample deformation. Therefore, the resolvability of real images should be determined not only from optical theory, but from the image itself. In crystallography, the resolution of each crystal is determined from the detection limit of high-angle diffraction in reciprocal space. In cryo-electron microscopy, correlation in the Fourier domain is used for estimating the resolution of sample images (Liao & Frank, 2010). We have evaluated the resolution of synchrotron radiation microtomography by using the slanted edge method (Mizutani et al., 2010a) and by using square-wave patterns (Mizutani et al., 2010b).

Crystal structures including those of proteins are determined from diffraction data. The detection limit of high-angle diffraction depends on atomic displacement parameters that represent the spatiotemporal displacements of atoms. The atomic displacement broadens the apparent distribution of the electron density of each atom, causing the high-angle diffractions to have lower intensities. This relationship can be illustrated as a linear correlation in a logarithmic intensity plot in reciprocal space (Wilson, 1942). The overall displacement parameter representing the average Gaussian broadening of electron densities of atoms can be estimated from this plot.

Fourier ring/shell correlation, which measures the normalized cross-correlation in the Fourier domain, has been used to estimate the resolution of electron microscopy images (Saxton & Baumeister, 1982; Haraus & van Heel, 1986). The resolution is estimated from the intersection of the Fourier ring/shell correlation and a threshold curve. Fourier ring correlation has also been applied to X-ray imaging microscopy (Vila-Comamala et al., 2012) and to X-ray ptychographic tomography (Holler et al., 2014).

Spatial resolution of X-ray images can be estimated by considering the noise level in the Fourier domain (Modregger et al., 2007), although there are a number of variations in defining the noise level criteria (Rose, 1973; Swank, 1974) and hence the resultant spatial resolution. In this paper, we report a method for estimating the spatial resolution from a logarithmic intensity plot in the Fourier domain without explicitly defining a noise criterion. The full width at half maximum (FWHM) of the Gaussian estimated from the plots of test images coincided with the FWHM of the Gaussian applied to the test images. The spatial resolution of imaging X-ray microtomography was estimated from the logarithmic intensity plot. We compared the result with the resolution determined from a micro-fabricated test object and with those estimated with other methods.

**Materials and methods**

*Test images*

A paraffin section of a zebrafish brain stained with Bodian impregnation (Mizutani et al., 2008b) was taken with a light microscope (Eclipse80i, Nikon) equipped with a CCD camera (DXM1200F, Nikon). The microphotograph was subjected to 4 × 4 binning in order to eliminate intrinsic blurring due to the microscope optics. A test chart was digitally clipped from the PM5544 test pattern (Philips). A black-white scan of texts was taken from the methods section of our previous report (Mizutani et al., 2010a). These images had dimensions of 500 × 500 pixels. A photograph of the experimental hutch of the BL37XU beamline of SPring-8 (Hyogo, Japan) was taken with a cell-phone CMOS camera (Sony Ericsson) and subjected to 2 × 2 binning, giving a 480 × 480 pixel image. The color images were converted into grayscale ones prior to the resolution estimation. Test images with various resolutions were generated by convolving these original images and a Gaussian point-spread function whose FWHM was 2, 4, 6, or 8 pixels.

*Tissue samples*

Human cerebral tissues were subjected to Golgi impregnation, as described previously (Mizutani et al., 2008a). The post-mortem brain tissues were collected with informed consent from the legal next of kin using protocols approved by ethical committees of the related organizations. The stained tissues were sequentially immersed in ethanol, *n*-butylglycidyl ether, and Petropoxy 154 epoxy resin (Burnham Petrographics). The samples were transferred to a borosilicate glass capillary (W. Müller) filled with the resin. The outer diameter of the capillary was 0.6-0.8 mm. The capillaries were kept at 90ºC for 40 hr to cure the resin.

*Test object*

An aluminum test object with square-wave patterns was prepared by using a focused ion-beam (FIB) apparatus (FB-2000; Hitachi High-Technologies) operated at 30 kV. An aluminum rod with approximate dimensions of 120 × 120 × 2000 μm was subjected to ion-beam milling. A gallium beam of 15 nA was used for rough abrasion of the aluminum surface, and a 0.5-2 nA beam was used to finish the flat surface. A series of square wells was carved on the surface. The pitches of the square-wave patterns were 2000, 1000, 500, 300, 240, 180, 140, 120, and 100 nm. Each pattern was composed of half-pitch wells and half-pitch intervals, *i.e.*, a 500-nm well and 500-nm interval for a 1000-nm pitch. A secondary electron image of the square-wave patterns is shown in Fig. 1. The test object was recovered and mounted on a stainless steel pin by using epoxy glue.

*Microtomography*

Imaging microtomography was performed at the BL37XU beamline (Suzuki et al., 2013a) of SPring-8. A Fresnel zone plate with an outermost zone width of 50 nm and diameter of 250 μm (Suzuki et al., 2005) was used as an X-ray objective lens, and an X-ray guide tube (Suzuki et al., 2013b) as a beam condenser. Transmission images produced by 8-keV X-rays were recorded using a CMOS-based X-ray imaging detector (AA60P and ORCA-Flash4.0, Hamamatsu Photonics) placed at 27.5 m from the sample. The detector viewing field was 2048 pixels in the horizontal direction perpendicular to the sample rotation axis and 2048 along the vertical axis. The effective pixel size was 23 nm. The resolution attained with this objective lens was examined using a Siemens star chart (XRESO-50HC, NTT Advanced Technology). Samples were fixed using brass fittings on a slide-guide rotation stage specially built for submicrometer resolution microtomography (Suzuki et al., 2013a; SPU-1A, Kohzu Precision). A total of 900 images per dataset were acquired with a rotation step of 0.20º and exposure time of 800 ms per image. The data collection conditions are summarized in Table 1.

The obtained transmission images were subjected to a convolution-back-projection calculation with the RecView program (Mizutani et al., 2010b; available from http://www.el.u-tokai.ac.jp/ryuta/). Each microtomographic slice perpendicular to the sample rotation axis was reconstructed with this calculation.

*Resolution estimation with logarithmic intensity plot*

Logarithmic intensity plots in the Fourier domain for the resolution estimation were calculated from the test images and also from the microtomographic slices. The structure $\rho(\mathbf{r})$ observed in the image can be expressed as the sum of the point spread functions (PSF) $\rho_i(\mathbf{r})$ of each structural constituent $i$,

$$\rho(\mathbf{r}) = \sum_i \rho_i(\mathbf{r}) = \sum_i \rho_i(\mathbf{r}_i + \mathbf{r}'),$$

where $\mathbf{r}_i$ is the coordinate of the center of PSF $\rho_i$, and $\mathbf{r}'$ the local coordinate within $\rho_i$. If we assume that each PSF $\rho_i$ with amplitude $a_i$ is proportional to the standard PSF $\rho_o(\mathbf{r}')$ with unit amplitude, each PSF can be written as

$$\rho_i(\mathbf{r}_i + \mathbf{r}') = a_i \rho_o(\mathbf{r}').$$

Then, the Fourier transform $F(\mathbf{k})$ of the structure $\rho(\mathbf{r})$ is calculated as

$$\begin{aligned} F(\mathbf{k}) &= \int \rho(\mathbf{r}) \exp(2\pi i \mathbf{r}\mathbf{k}) dv \\ &= \sum_i \int \rho_i(\mathbf{r}_i + \mathbf{r}') \exp[2\pi i (\mathbf{r}_i + \mathbf{r}')\mathbf{k}] dv \\ &= \sum_i a_i \exp(2\pi i \mathbf{r}_i \mathbf{k}) \int \rho_o(\mathbf{r}') \exp(2\pi i \mathbf{r}'\mathbf{k}) dv'. \end{aligned}$$

Here, we define the Fourier transform of the standard PSF as

$$f_o(\mathbf{k}) = \int \rho_o(\mathbf{r}') \exp(2\pi i \mathbf{r}'\mathbf{k}) dv'.$$

The square norm of $F(\mathbf{k})$ is

$$\begin{aligned} |F(\mathbf{k})|^2 &= F(\mathbf{k}) F^*(\mathbf{k}) \\ &= \sum_i a_i \exp(2\pi i \mathbf{r}_i \mathbf{k}) f_o(\mathbf{k}) \cdot \sum_j a_j \exp(-2\pi i \mathbf{r}_j \mathbf{k}) f_o^*(\mathbf{k}) \\ &= \sum_i \sum_j a_i a_j f_o(\mathbf{k}) f_o^*(\mathbf{k}) \exp[2\pi i (\mathbf{r}_i - \mathbf{r}_j)\mathbf{k}] \\ &= |f_o(\mathbf{k})|^2 \left[ \sum_i a_i^2 + \sum_i \sum_{j \neq i} a_i a_j \exp[2\pi i (\mathbf{r}_i - \mathbf{r}_j)\mathbf{k}] \right]. \end{aligned}$$

If the positional relationship ($\mathbf{r}_i$ - $\mathbf{r}_j$) can be regarded as being random and if each $a_i$ has comparable amplitude (corresponding to a homogeneous random structure), the second term in the parentheses can be ignored. Therefore, the square norm logarithm can be approximated with

$$\ln|F(\mathbf{k})|^2 \cong 2\ln|f_o(\mathbf{k})| + \ln\left(\sum_i a_i^2\right).$$

The PSF constituting the image should be expressed with a certain function for the resolution estimation. The PSF of the real image depends not only on the optical system but also on the mechanical drift and sample deformation. Though it is difficult to represent all these blurring effects with a simple model function, a PSF approximated with a Gaussian can be easily identified by drawing a logarithmic plot in the Fourier domain. Therefore, we approximated the standard PSF with a Gaussian:

$$\rho_o(\mathbf{r}') = \exp\left(-\frac{|\mathbf{r}'|^2}{2\sigma^2}\right).$$

This assumption was proven appropriate in the resolution estimation of a sample image, as described below. The Fourier transform of this standard PSF in the *n*-dimensional space is expressed as

$$f_o(\mathbf{k}) = \left(\sqrt{2\pi}\sigma\right)^n \exp\left(-2\pi^2\sigma^2|\mathbf{k}|^2\right).$$

This gives us the relationship,

$$\ln|F(\mathbf{k})|^2 \cong -4\pi^2\sigma^2|\mathbf{k}|^2 + \ln\left(\sum_i a_i^2\right)(2\pi\sigma^2)^n,$$

indicating that the width $\sigma$ of the standard PSF $\rho_o(\mathbf{r}')$ can be determined by plotting $\ln|F(\mathbf{k})|^2$ as a function of $|\mathbf{k}|^2$. A similar plot in crystallography is known as the Wilson plot (Wilson, 1942).

Figure 2 illustrates the procedure of estimating the spatial resolution from the logarithmic intensity plot in the Fourier domain. First, each image was subjected to a Fourier transform giving a frequency-domain pattern (Fig. 2B). The Fourier transform calculations were performed with the RecView program. The logarithm of the average squared norm in each 5 × 5 pixel bin of the Fourier transform was plotted against the squared distance from the origin (Fig. 2C). The region along the reciprocal coordinate axes was not included in the plot. The spatial resolution was then estimated from the linear correlation representing the FWHM of point spread function. Since this estimate derives from the Fourier transform of the image, it represents the resolution of the entire image and does not represent the resolution of a certain local structure in the image.

*Fourier ring correlation*

Fourier ring/shell correlation measures the normalized cross-correlation in the Fourier domain as a function of spatial frequency:

$$\text{Fourier ring/shell correlation}(k) = \frac{\sum_{i \in k} F_i G_i^*}{\sqrt{\left(\sum_{i \in k} |F_i|^2\right)\left(\sum_{i \in k} |G_i|^2\right)}},$$

where the correlation in shell $k$ is calculated from the Fourier transforms $F_i$ and $G_i$ of a pair of images. The correlation values were calculated using the FSC program of the IMAGIC software suite (Van Heel et al., 1996; Image Science Software GmbH, Berlin). The size factor (Van Heel & Schatz, 2005) was not used in this study. In the resolution estimation, the tomographic dataset was divided into odd and even numbered frames, from which independent tomographic images of the same slice were reconstructed. Then the Fourier ring correlation was calculated from the central 300 × 300 pixels of the tomographic images in order to keep the angular sampling step finer than the Nyquist limit.

**Results and discussion**

*Resolution estimation using test images*

Test images generated by convolving the Gaussian point-spread function and the original image (Fig. 3A-D) were analyzed with the logarithmic intensity plot in the Fourier domain. The appearance of square wave patterns embedded in the original image proved that the Gaussian function applied to the original image introduced blurring corresponding to the Gaussian

FWHM. Logarithmic intensity plots of the test images are shown in Fig. 3E-H. Linear correlations corresponding to the Gaussian were observed at the left ends of these plots. These indicated that the Gaussian point-spread function can be extracted from the image. The FWHMs of the point spread function were calculated from the slopes. For example, the slope of the plot in the left half of Fig. 3G was $\Delta\ln|F(\mathbf{k})|^2 / \Delta|\mathbf{k}|^2 = -316$ (pixel$^2$). The standard deviation of the Gaussian corresponding to this slope was calculated to be $\sigma = \sqrt{-\text{slope}/2\pi}$ = 2.8 pixels. This gave a FWHM of $2.8 \times 2\sqrt{2\ln 2}$ = 6.6 pixels, which is comparable to the FWHM (6 pixels) of the Gaussian applied to the original image and also consistent with the appearance of square-wave patterns embedded in the test image. Therefore, the obtained FWHM can be regarded as a measure of image resolvability.

The logarithmic intensity plot in the Fourier domain estimates the FWHM of the point spread function without defining any noise level criterion. Noise in the original image rather is observed as a noise profile in the plot. Figure 4 shows the effect of noise on this method. Pixel-by-pixel uniform noise with amplitudes ranging from -10% to 10% of the average amplitude of the original image was added to the image shown in Fig. 3B. The obtained plot (Fig. 4B) showed a shallow slope at the right end, supposedly representing the noise, while the slope of the Gaussian point-spread function was separately observed in the left half of the plot. These results indicated that the obtained PSF represents the resolution of the sample object, of which the signal is larger than the noise level. Since the resolution can be estimated if the sample slope is separated from the baseline profile, noise level criteria need not be defined in this method.

Spatial resolutions of several test images (Fig. 5) were also estimated with this method. The linear correlation was identified independently of the type of original image, including the PM5544 test pattern shown in Fig. 5E. We suggest that the homogeneous randomness assumed in the Materials and methods section is not a stringent prerequisite of this method. The Nyquist limit of the digitization of the original image is the lower limit of the FWHM estimation. Therefore, the estimated FWHM should be $[(\text{applied FWHM})^2 + w_N^2]^{1/2}$, where $w_N$ represents the Nyquist limit. The value $w_N$ is 2 pixels, since structures only down to twice the sampling width are resolvable in the digitized image. The FWHM of the point spread function estimated from the logarithmic intensity plot almost coincided with this prediction (Fig. 5G), indicating that the spatial resolution of any type of image can be estimated with this method.

*Resolution estimation of microtomographic slices*

The highest possible resolution of the imaging microtomography used in this study was expected to be 100-150 nm. Square-wave patterns with pitches corresponding to this resolution were prepared on the aluminum surface. A back-scattered electron image of the test object is shown in Fig. 1; the patterns had pitches as narrow as 100 nm. The minimum well width that could be carved with the FIB apparatus was approximately 50 nm, and this made it hard to make square-wave patterns with pitches less than 100 nm.

Figure 6A shows a microtomographic cross section of the test object. The cross section indicated that the patterns up to 120 nm pitches were clearly resolved. Since some structures of the 100-nm pitch pattern were visualized, we suggest that the resolution of this microtomograph is about 100-120 nm. The innermost 50-nm line and space of a Siemens star chart were resolved in a still X-ray image (Fig. 6B), indicating that the spatial resolution achieved with the X-ray optics is finer than 100 nm. Therefore, the tomographic resolution should be limited by the stability of the sample during the dataset collection, which depends on the wobbling of the sample rotation stage.

Figure 7A shows a microtomographic cross section of a human pyramidal neuron taken with the imaging microtomography. Cellular nucleus, dendrite, and cytosolic components are visualized in this cross section. The spatial resolution of this cross section was estimated from the logarithmic intensity plot in the Fourier domain. A linear correlation should be observed if the intensity distribution can be considered random and if the point spread function constituting this image can be approximated with a Gaussian. Although the neuronal structure is not random but rather organized, the cytosolic constituents exhibited randomness. The logarithmic intensity plot (Fig. 7B) was calculated from the boxed area (Fig. 7A) located at the sample rotation axis. Here, a linear correlation was observed at the left end. The FWHM of the Gaussian estimated from this plot was 118 nm, which coincides with the resolution (100-120 nm) determined from the test object.

The shallow slope in the right half of Fig. 7B represents a spatial frequency of twice the pixel width. This corresponds to the Nyquist limit of digitization. A similar profile was also seen in the noise simulation (Fig. 4B). Therefore, this slope should originate from pixel-by-pixel noise, such as detector dark current or tomographic reconstruction artifacts, and should not be used for the resolution estimation. Fourier profiles composed only from the slope representing the Nyquist limit were found in some experiments. Such profile indicates that the image resolvability solely rests on the detector pixel width and that the spatial resolution achieved by the optics is finer than the Nyquist limit of image digitization.

*Comparison with other methods*

Spatial resolution of X-ray images has been estimated by considering noise level in the Fourier domain (Modregger et al., 2007). It has been proposed that the effective resolution limit can be estimated from the maximum spatial frequency at which the Fourier power spectrum surpasses twice the noise level. Figure 8A shows an amplitude plot of the Fourier transform of the boxed area in Fig. 7A. The flat profile at the high-frequency end can be regarded as the noise level. The dashed line in Fig. 8A corresponding to √2-times this level in amplitude (twice the level in power spectrum) intersects the plot at a spatial frequency of 1/60 nm$^{-1}$. This is nearly twice as high as that estimated from the logarithmic plot of the same image (1/118 nm$^{-1}$), and that determined using the test object (1/100 - 1/120 nm$^{-1}$).

The Rose criterion states that the signal amplitude should be five times larger than the root-mean-square noise for reliable observation (Rose, 1973). The threshold of five times the noise level (Fig. 8A) intersects the Fourier domain plot at 1/90 nm$^{-1}$, which coincided with the test object image rather well. Though the Rose criterion was developed for real domain analysis, this result suggests that it should also be applied in the Fourier domain.

Fourier ring/shell correlation, which measures the normalized cross-correlation in the Fourier domain, has been used to estimate the resolution of electron microscopy images (Saxton & Baumeister, 1982; Haraus & van Heel, 1986). The resolution is estimated from the intersection of the Fourier ring/shell correlation and a threshold curve, such as the half-bit information curve and 3-sigma curve (Van Heel & Schatz, 2005). The Fourier ring correlation calculated from the neuron image is shown in Fig. 8B. The resolution was estimated from the point where the Fourier ring correlation intersects the half-bit curve or 3-sigma curve. The resolution estimated from the half-bit curve was 170 nm, and that from the 3-sigma curve was 140 nm, both coarser than the resolution determined from the test object. The resolution estimation with this method depends on threshold curves that intersect the sample profile at small angles. A slight shift of the threshold curve or sample profile can alter the intersection position, resulting in variations in the estimated resolution.

**Conclusion**

The microtomographic cross section of the test object indicated that a spatial resolution as fine as 120 nm can be determined using square-wave patterns. This resolution value is not determined freely but from pattern pitches. The fabrication precision of the FIB apparatus also poses a limitation on the minimum pitch of the square-wave pattern (Mizutani et al., 2010a). However, the visualization of the pattern clearly illustrated the achieved resolution without any ambiguity. The resolution should therefore be evaluated by using square-wave patterns if patterns with pitches of the target resolution are available.

If such a test object cannot be prepared or if the image variation should be included in the resolution, the spatial resolution has to be estimated from the sample image itself. In this study, we reported a resolution estimation method that uses a logarithmic intensity plot in the Fourier domain. This method can provide a resolution measure that is directly related to the FWHM of the point spread function. The advantage of this method is that the resolution can be estimated without explicitly defining a noise criterion. Since the estimate derives from the Fourier transform of the image, it represents the resolution of the entire image and not the resolution of a certain local structure.

The resolution of the neuron image (118 nm) estimated with the logarithmic intensity plot in the Fourier domain showed coincidences with the test object resolution (100-120 nm). The Fourier ring correlation underestimated the spatial resolution of the neuron image. The 2-sigma noise criterion in a Fourier domain plot overestimated the spatial resolution, while the Rose criterion gave an estimate rather comparable to the test object resolution. The variations (60-170 nm) in these criterion-based estimates are ascribable to the definition of the threshold criterion. In contrast, the logarithmic intensity plot provided a sound estimate without explicitly defining any criterion. We suggest that the logarithmic intensity plot in the Fourier domain provides an alternative measure of image resolvability regardless of visualization modality.

**Acknowledgements**

We thank Yasuo Miyamoto and Kiyoshi Hiraga (Technical Service Coordination Office, Tokai University) for helpful assistance with FIB milling and preparation of brass fittings for microtomography. We thank Noboru Kawabe (Teaching and Research Support Center, Tokai University School of Medicine) for helpful assistance with the histology. This study was supported in part by Grants-in-Aid for Scientific Research from the Japan Society for the Promotion of Science (nos. 25282250 and 25610126). The synchrotron radiation experiments were performed at SPring-8 with the approval of the Japan Synchrotron Radiation Research Institute (JASRI) (proposal nos. 2013A1384, 2013B0041, and 2014A1057).

**Figure captions**

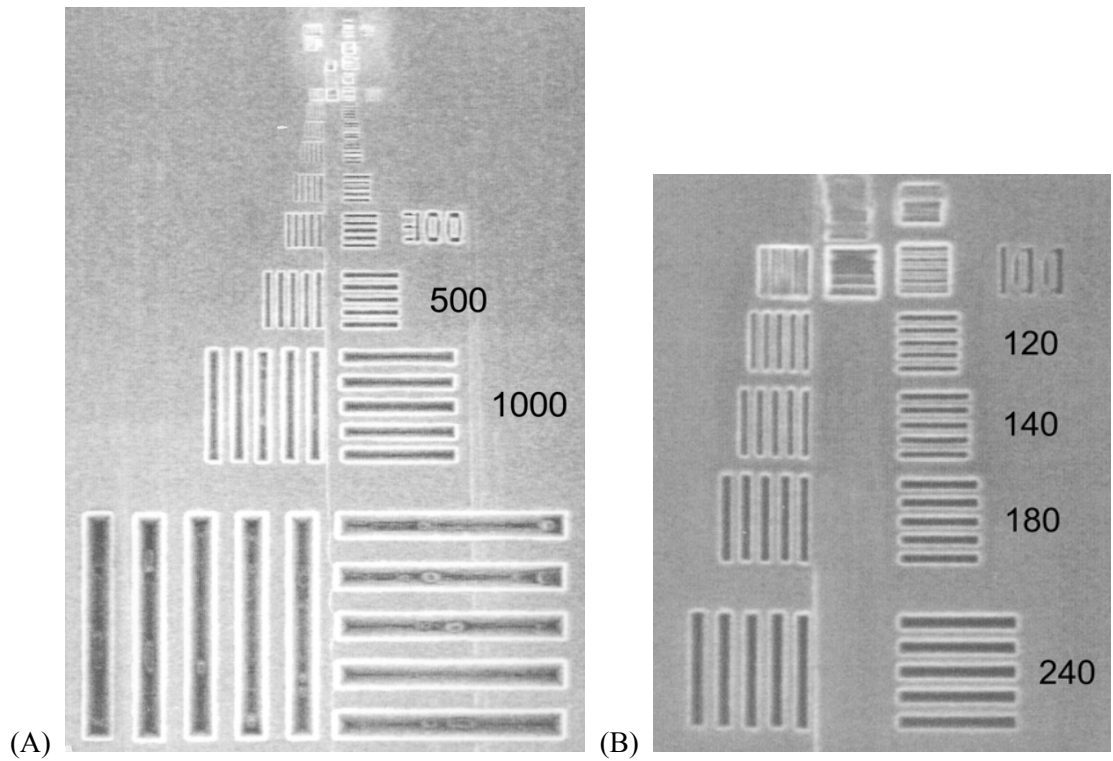

(A)  (B)

Figure 1. Secondary electron images of aluminum test object. (A) Overall view of the entire patterns. Patterns with pitches of 500 and 1000 nm are indicated with labels. The 300-nm pattern has its pitch carved next to it. (B) Magnified image of square-wave patterns with pitches of 240, 180, 140, 120 and 100 nm. The 100-nm pattern has its pitch carved next to it. Other patterns are indicated with labels.

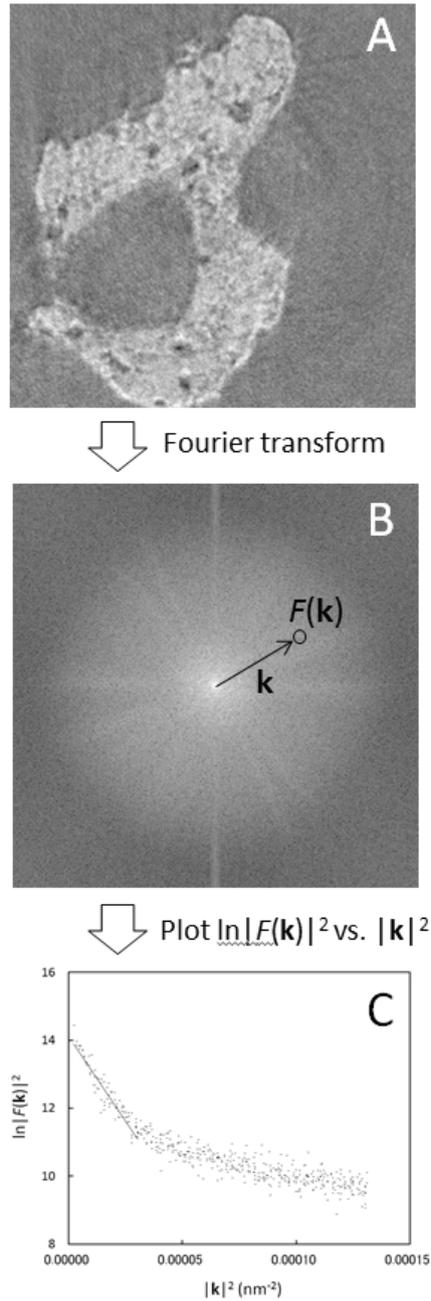

Figure 2. Estimation of spatial resolution from logarithmic intensity plot in the Fourier domain. First, original image (A) was subjected to Fourier transform, giving a frequency-domain pattern (B). Then, the logarithm of squared norm of the Fourier transform was plotted against squared distance from the origin (C). Spatial resolution was estimated from the linear correlation observed at the left end of the plot.

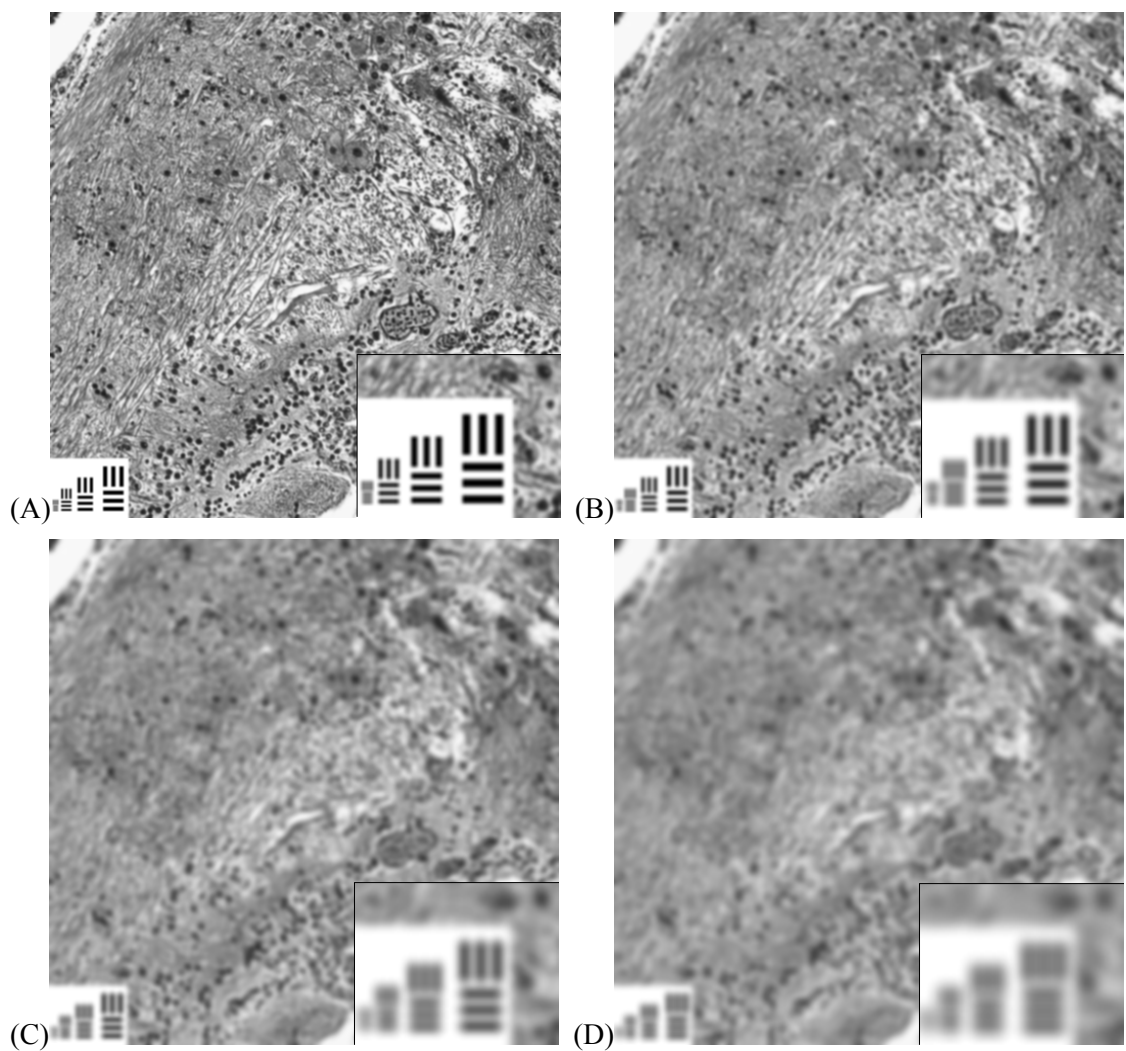

Figure 3. Test images of a paraffin section of zebrafish brain were generated by convolving the original image with Gaussian point-spread function whose FWHM was 2 (A), 4 (B), 6 (C), or 8 pixels (D). The lower right inset in each panel shows a two-fold magnification of square-wave patterns with pitches of 2, 4, 6, and 8 pixels.

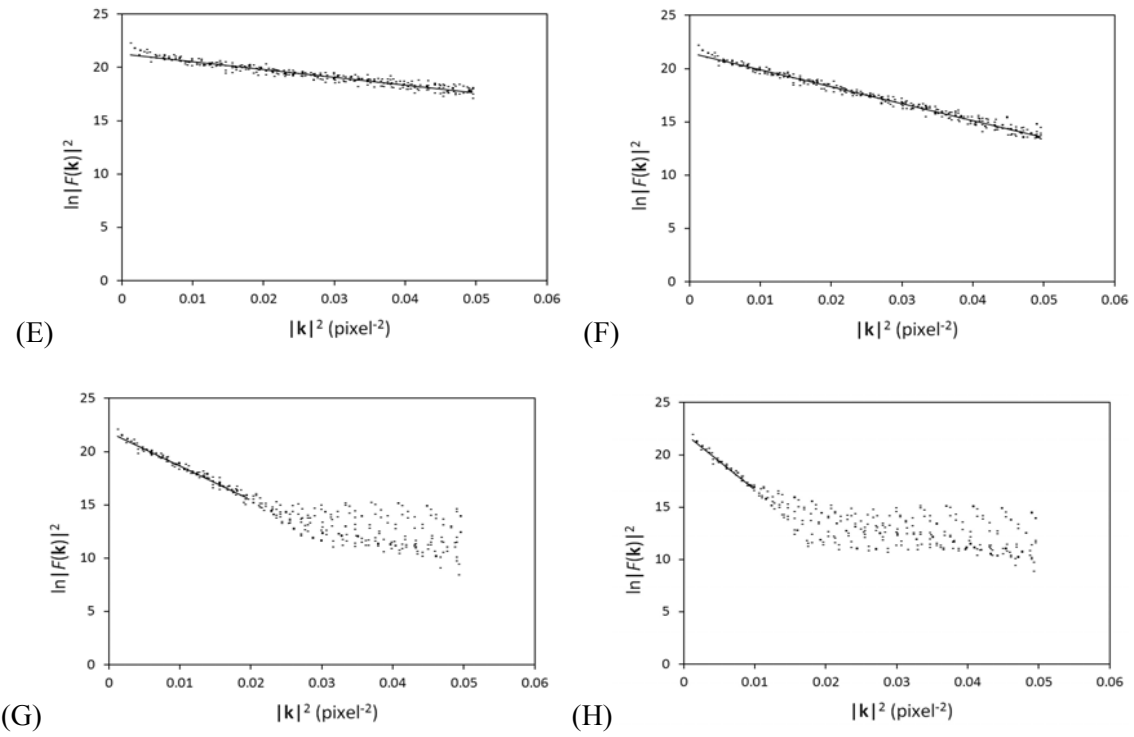

Figure 3. Panels E-H show logarithmic intensity plots calculated from the test images shown in A-D, respectively. Linear correlations representing Gaussian point-spread functions are indicated with thin lines.

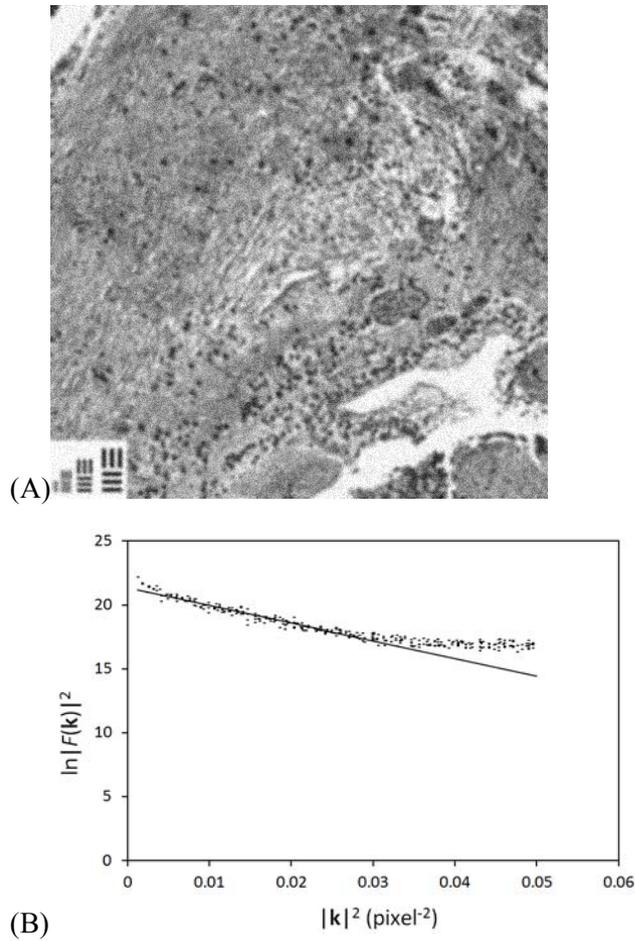

Figure 4. (A) Pixel-by-pixel noise with amplitudes ranging from -10% to 10% of the original image average was added to the image shown in Fig. 3B. (B) The resultant image was subjected to the logarithmic intensity plot. The noise profile was observed at the right end, while the slope corresponding to the Gaussian FWHM (4 pixels) was observed in the left half.

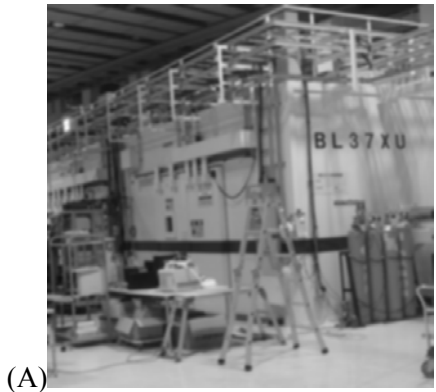
(A)
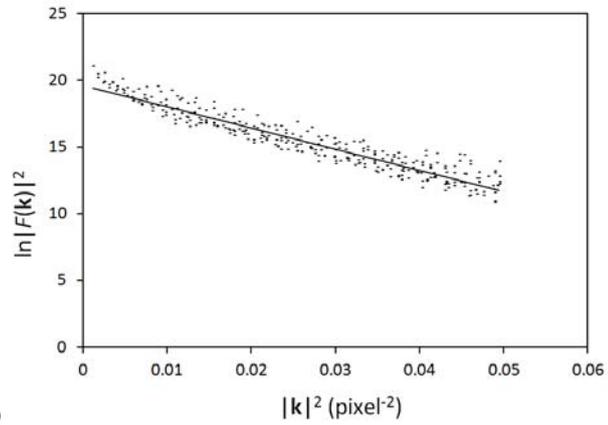
(B)

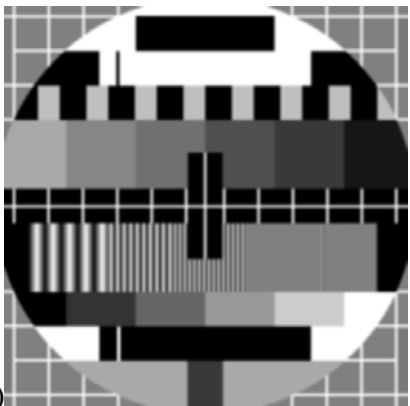
(C)
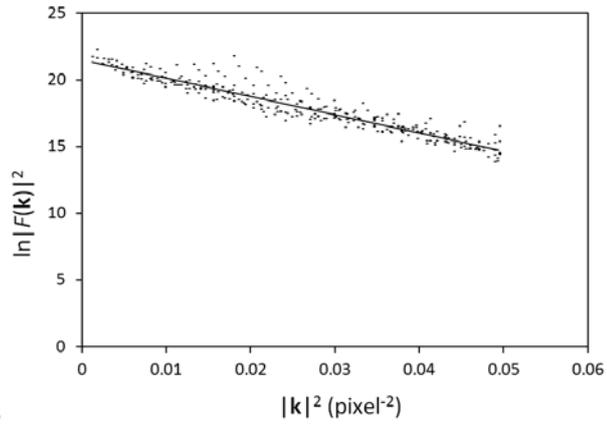
(D)

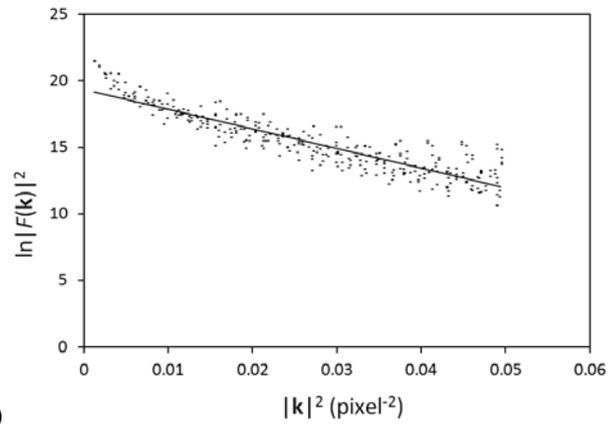
(E)
(F)

Figure 5. The photograph of the BL37XU hutch of SPring-8 (A), black-white scan of texts (C), and grayscale test pattern (E) were generated by convolving the original images with a Gaussian having an FWHM of 4 pixels. Logarithmic intensity plots (B, D, and F) were calculated using the images in panels A, C, and E, respectively. Linear correlations representing Gaussian point-spread functions are indicated with thin lines.

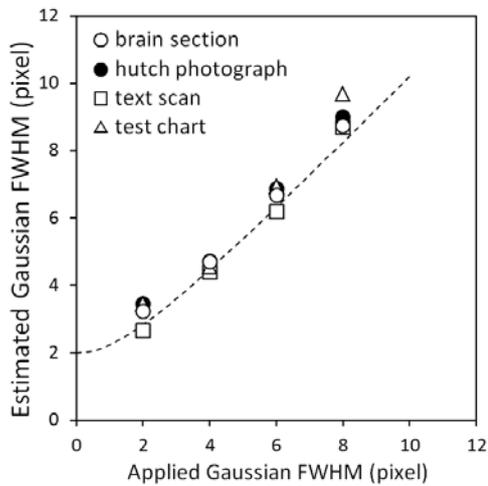

(G)

Figure 5. (G) Relationship between FWHMs of the Gaussians estimated from the logarithmic intensity plot and those of the Gaussians applied to the original image. Plots for the brain section image are indicated with open circles, hutch photograph with closed circles, text scan image with squares, and PM5544 test chart with triangles. The dashed line indicates the prediction: $[(\text{Applied FWHM})^2 + 2^2]^{1/2}$.

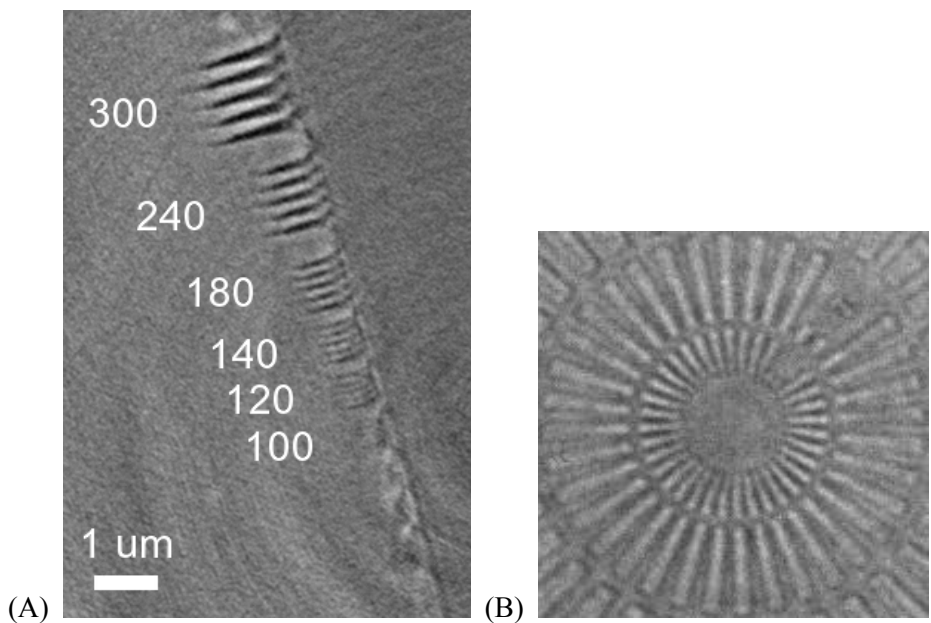

(A)  (B)

Figure 6. (A) Microtomographic cross section of square-wave patterns with pitches of 300, 240, 180, 140, 120 and 100 nm. Patterns with pitches up to 120 nm were resolved. Scale bar: 1 μm. (B) Transmission image of a star chart. The innermost 50-nm line and space were resolved.

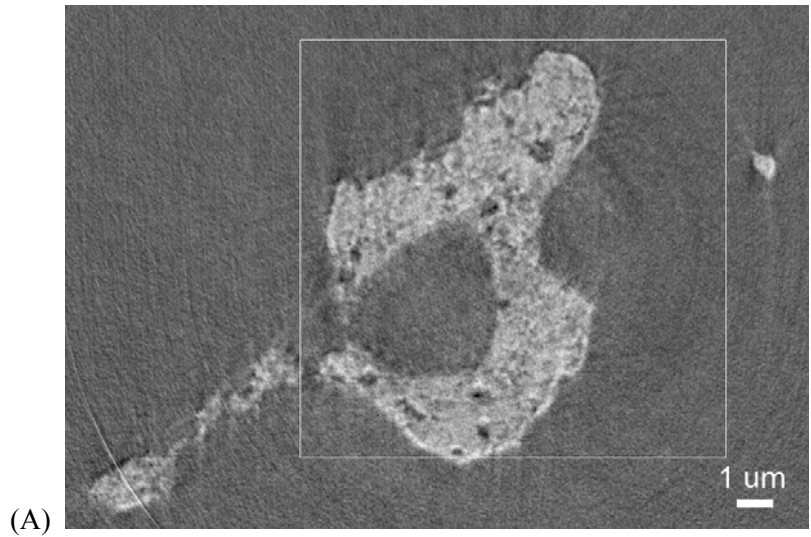

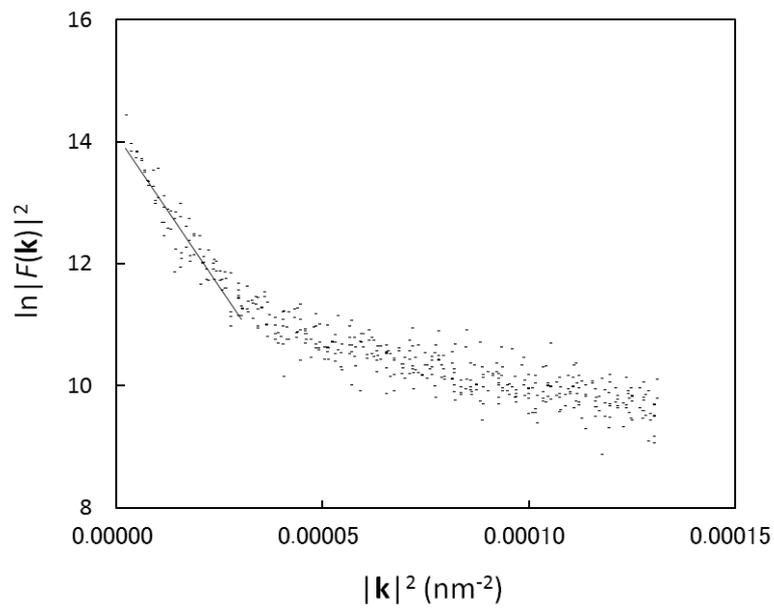

Figure 7. (A) Microtomographic cross section of a human pyramidal neuron. Spatial resolution was estimated from the 500 × 500 pixel area indicated by the box. Scale bar: 1 μm. (B) Logarithmic intensity plot of the boxed area. The linear correlation at the left end can be regarded as representing a Gaussian point-spread function.

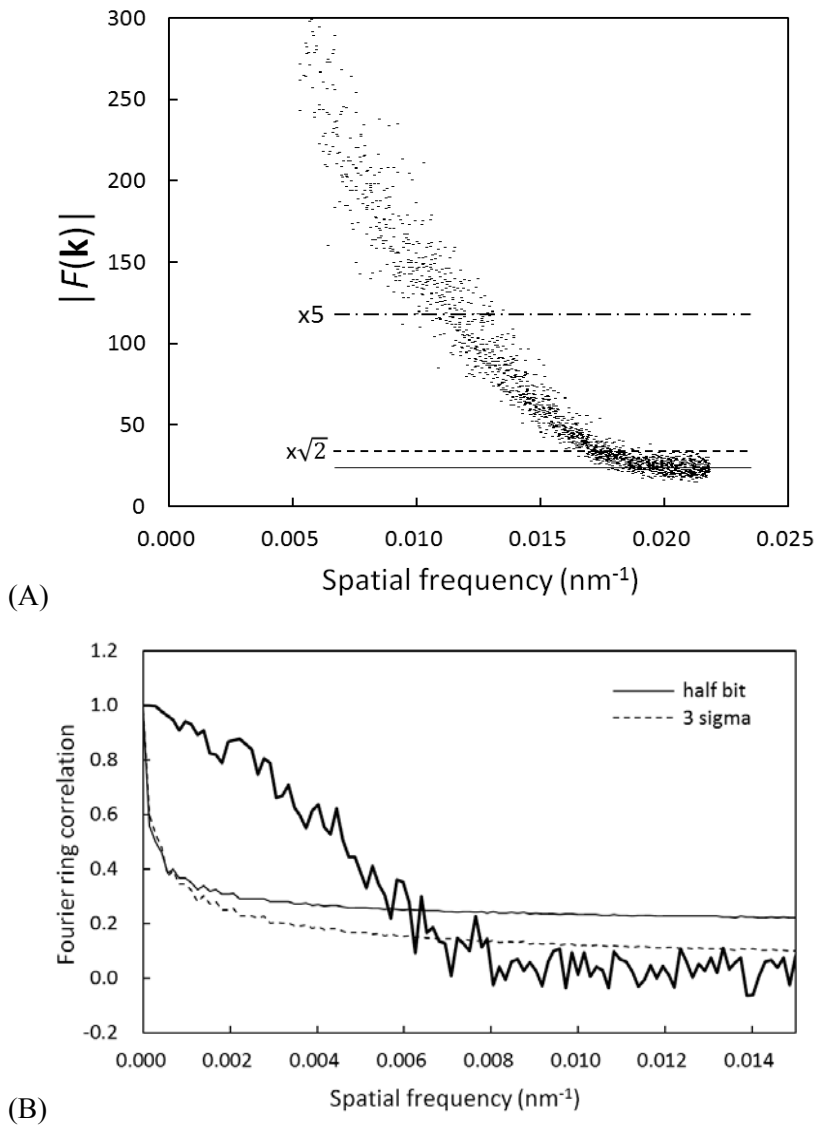

(A)

(B)

Figure 8. (A) Fourier amplitude plot of the neuron image shown in Fig. 7A. A thin line represents a possible noise level. Threshold criteria corresponding to √2-times (dashed line) and 5-times (dot-dash line) the noise level are also indicated. The upper part ($|F(\mathbf{k})| > 300$) of the plot is not shown. (B) Resolution estimation from Fourier ring correlation (bold line) of the neuron image. The half-bit information-threshold curve is indicated with a thin line, and the 3-sigma curve with a dashed line.

**Table 1.** Data collection conditions.

| Beamline | BL37XU |
| --- | --- |
| X-ray optics | Fresnel zone plate optics |
| X-ray energy (keV) | 8.0 |
| Pixel size (nm) [a] | 23 × 23 |
| Viewing field size (pixels) [a] | 2048 × 2048 |
| Viewing field size (μm) [a] | 47 × 47 |
| Rotation/frame (degrees) | 0.20 |
| Exposure/frame (ms) | 800 |
| Frame/dataset | 900 |
| Dataset collection time (min) | 15 |

[a] Width × height